\documentclass[11pt]{article}
\usepackage{moriond,epsfig}

\def\Journal#1#2#3#4{{#1} {\bf #2}, #3 (#4)}

\def\PRL{\em Phys. Rev. Lett.}
\def\PRD{{\em Phys. Rev.} D}

\def\be{\begin{equation}}
\def\ee{\end{equation}}
\def\bea{\begin{eqnarray}}
\def\eea{\end{eqnarray}}

\newcommand{\kslash}{k\kern-1ex /}
\newcommand{\pslash}{p\kern-1ex /}
\newcommand{\qslash}{q\kern-1ex /}
\newcommand{\lslash}{l\kern-1ex /}
\newcommand{\sslash}{s\kern-1ex /}
\newcommand{\Dslash}{D\kern-1.2ex /}

\newcommand{\beqa}{\begin{eqnarray}}
\newcommand{\eeqa}{\end{eqnarray}}
\newcommand{\bd}{\begin{description}}
\newcommand{\ed}{\end{description}}
\newcommand{\ben}{\begin{eqnarray}}
\newcommand{\een}{\end{eqnarray}}

\def\lsim{\raise0.3ex\hbox{$<$\kern-0.75em\raise-1.1ex\hbox{$\sim$}}}
\def\gsim{\raise0.3ex\hbox{$>$\kern-0.75em\raise-1.1ex\hbox{$\sim$}}}
\def\simgt{\rlap{\lower 3.0 pt\hbox{$\mathchar \sim$}}\raise 1.5pt \hbox {$>$}}
\def\simlt{\rlap{\lower 3.0 pt\hbox{$\mathchar \sim$}}\raise 1.5pt \hbox {$<$}}

\begin{document}
\vspace*{4cm}
\title{CURRENT STATUS TOWARD THE PROTON MASS CALCULATION\\ IN LATTICE QCD}

\author{Yoshinobu Kuramashi}

\address{Graduate School of Pure and Applied Sciences
and  Center for Computational Sciences,\\ 
University of Tsukuba, Tsukuba, Ibaraki 305-8571, Japan}

\maketitle\abstracts{
The proton mass calculation is still a tough challenge for lattice
QCD. We discuss the current status and difficulties
based on the recent PACS-CS results for
the hadron spectrum in 2+1 flavor QCD.}

\section{Introduction}

The proton mass calculation has a profound meaning in lattice QCD:
To distinguish the proton mass from the neutron one we need to incorporate 
the isospin breaking effects with the different up and down quark
masses and the electromagnetic interactions.
This is still out of reach for current lattice QCD calculations.
The accomplishment of the first principle calculation of the proton mass
inevitably means that other physical quantities should be calculated 
with similar precision on the same configurations.  

In this report we show the recent progress in lattice QCD based on the 2+1
flavor lattice QCD results obtained by 
the PACS-CS Collaboration who are currently aiming at the physical
point simulation.\cite{pacscs_nf3}
We discuss the difficulties in lattice QCD from a view point of the
systematic errors. The toughest problem is the rapid increase of 
computational cost with the up-down (ud) quark mass reduced toward the
physical value.  
We explain why the direct simulation on the physical point is 
required in order to avoid the problems 
in the chiral extrapolation method. 

\section{Difficulties in Lattice QCD Calculation}

Most fundamental quantities in lattice QCD 
are Green functions in the path-integral formalism:
\ben
\langle {\cal O}[U,q,{\bar q}]\rangle &=& \frac{1}{Z}\int {\cal D}U{\cal D}q{\cal D}{\bar q}
{\cal O}[U,q,{\bar q}]{\rm e}^{-S_{\rm QCD}^{\rm L}[U,q,{\bar q}]},
\een   
where $S_{\rm QCD}^{\rm L}$ represents the QCD action 
defined on the discretized
four-dimensional space time. $U$ is the so-called link variable which
contains the gauge fields. $q$ and ${\bar q}$ denote 
the quark and the anti-quark fields.
Only the Monte Carlo method makes feasible the nonperturbative evaluation 
of the above expression.  
With appropriate choices of the operator ${\cal O}$ we can extract 
various physical quantities, {\it e.g.}, the hadron spectrum.

\begin{figure*}[h]
  \begin{tabular}{lr}
    \begin{minipage}[c]{0.45\textwidth}
\psfig{figure=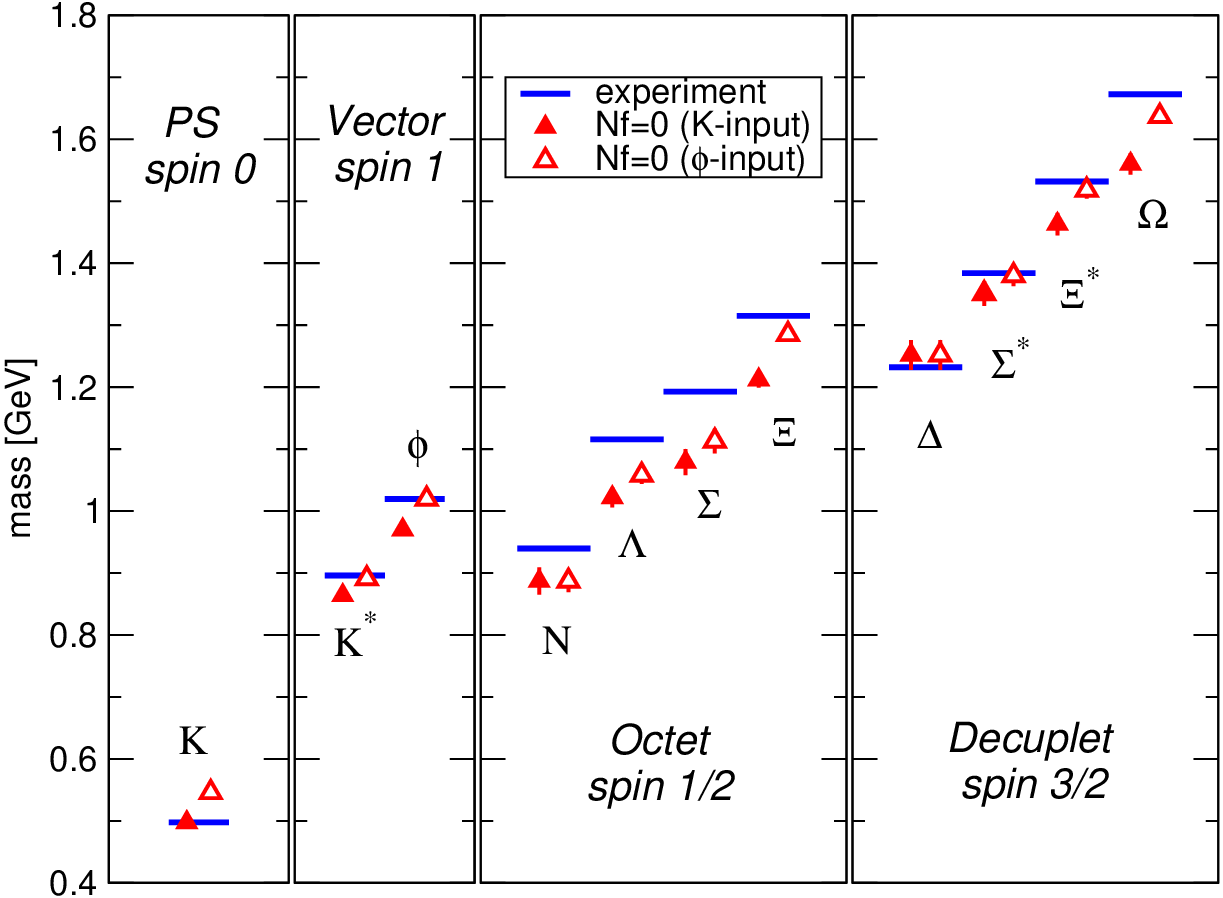,angle=0,height=48mm}
      \caption{Quenched light hadron spectrum compared with
experiment.}
      \label{fig:cppacs_nf0}
    \end{minipage}
&
    \begin{minipage}[c]{0.45\textwidth}
\psfig{figure=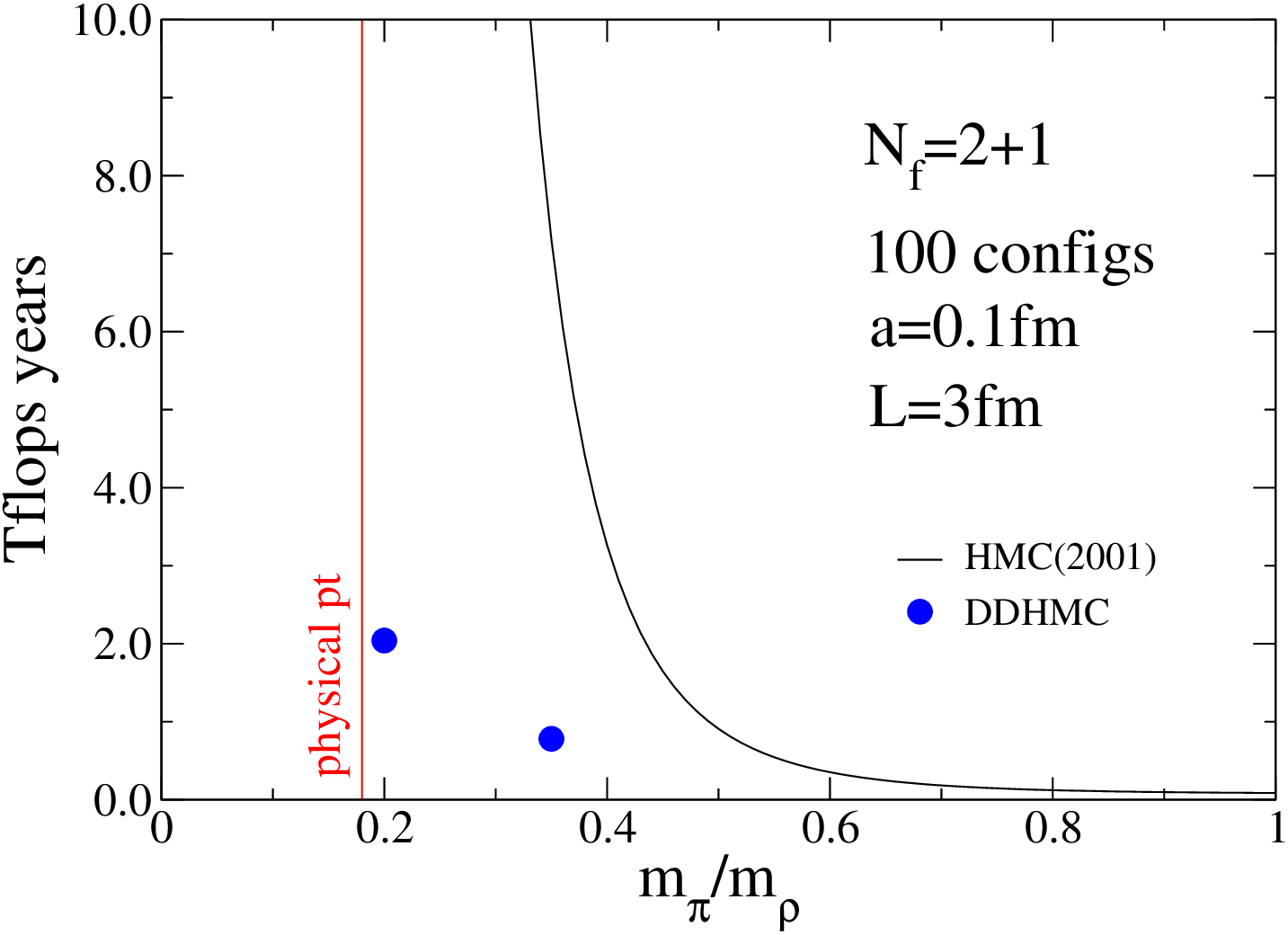,angle=0,height=52mm}
      \caption{Simulation cost as a function of $m_\pi/m_\rho$. See
text for details.}
      \label{fig:cost}
    \end{minipage}
  \end{tabular}
\end{figure*}

There exists two types of errors in lattice QCD: 
One is the statistical one due to the Monte Carlo technique. 
The other is the systematic ones.
The former is arbitrarily reduced according to $1/\sqrt{N}$ with $N$ the
number of the independent configurations (Monte Carlo samples).
The troublesome is the latter.
We have four major systematic errors: (i) finite
volume effects, (ii) finite lattice spacing effects, (iii) quench
approximation and (iv) chiral extrapolation. 
It is rather straightforward to diminish 
the first and the second errors with the use of larger and finer lattices.
For almost twenty years after the first lattice QCD calculation 
of the hadron masses 
in 1981,\cite{Parisi} most of the large-scale simulations were
carried out in the quenched approximation where the
sea quark effects are neglected. The primary reason is that 
the 2+1 flavor lattice QCD simulation
requires $O(10^2)$ times as much computational cost 
as the quenched approximation.
In late 90s the CP-PACS collaboration performed 
a detailed investigation of the quenching effects.\cite{cppacs_nf0}  
The systematic study of the hadron spectrum in the quenched
approximation with other systematic errors under control
reveals that the results deviate from the experimental values at a 10\% level.
The comparison are depicted in Fig.~\ref{fig:cppacs_nf0}, where the
physical inputs are a set of $m_\pi, m_\rho, m_K$ (closed triangles)
or $m_\pi, m_\rho, m_\phi$ (open triangles) to
determine the averaged up-down quark mass, the strange one and the
lattice spacing $a$. The confirmation of the discrepancy between the
quenched results and the experimental values drove us to embark on
the 2+1 flavor QCD simulations.\\
\vspace*{1mm}   

Now the remaining task is to remove 
the systematic error associated with the chiral extrapolation.
Figure~\ref{fig:cost} illustrates the difficulty: The
solid line represents the empirical cost estimate for
the 2+1 flavor lattice QCD simulation with the Hybrid Monte Carlo (HMC)
algorithm given by Ukawa in 2001.\cite{berlinwall} 
The cost seems to almost diverge as the $m_\pi/m_\rho$ ratio 
approaches the physical point. It was obvious that we definitely need
not only the increase of the computational power but also the
algorithmic improvements. Years later the difficulty is overcome by
the Domain-Decomposed Hybrid Monte Carlo (DDHMC)
algorithm.\cite{ddhmc} Blue circles denote the measured computational
cost in our simulation armored with several other algorithmic
improvements, which clearly shows that the direct simulation at the
physical point is allowed with the current computational resources.

\begin{figure*}[h]
  \begin{tabular}{lr}
    \begin{minipage}[t]{0.45\textwidth}
\psfig{figure=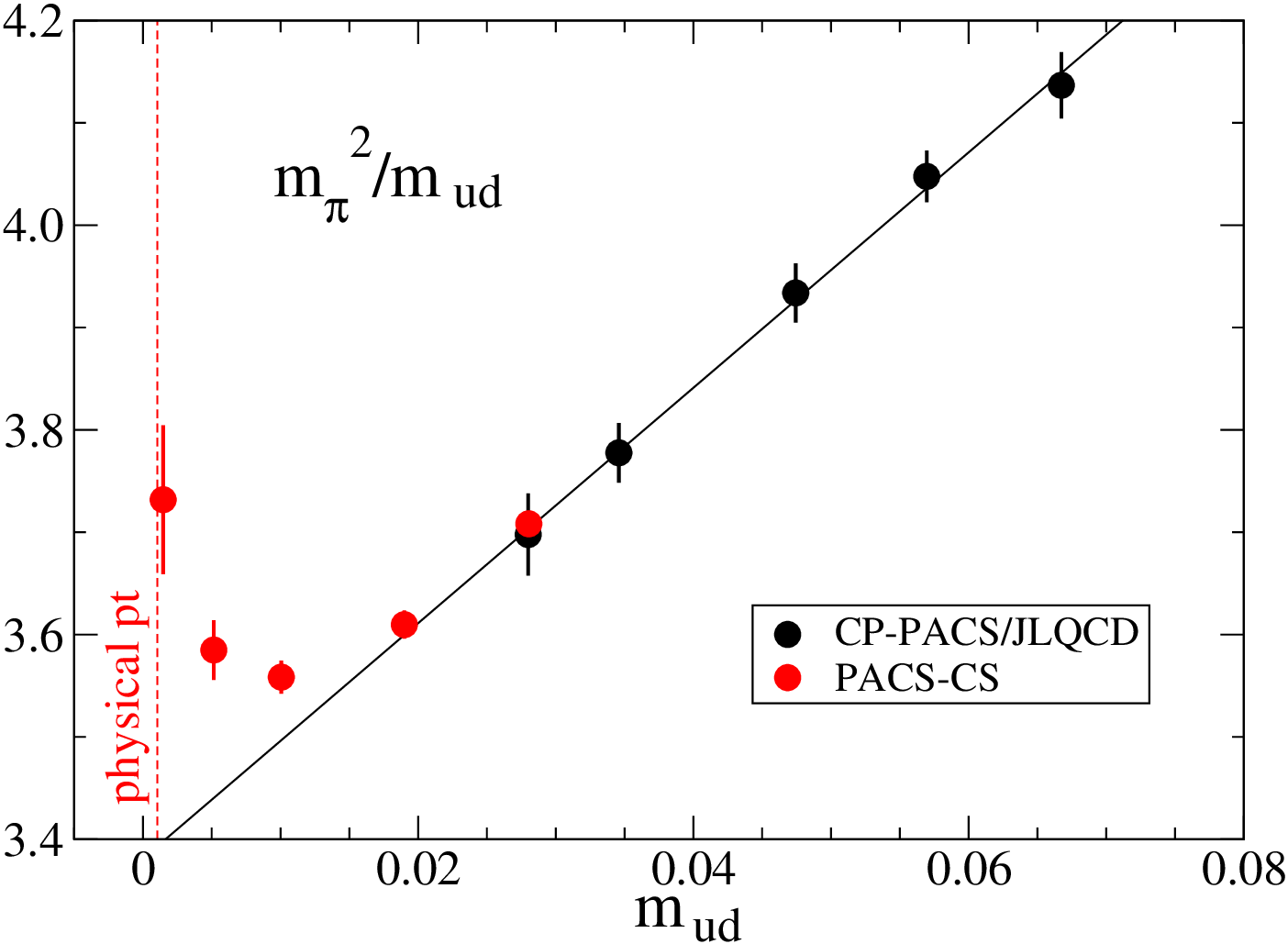,angle=0,height=50mm}
      \caption{$m_\pi^2/m_{\rm ud}$ as a function of $m_{\rm ud}$.
Solid line is just for guiding your eyes.}
      \label{fig:mpi2mud}
    \end{minipage}
&
    \begin{minipage}[t]{0.45\textwidth}
\psfig{figure=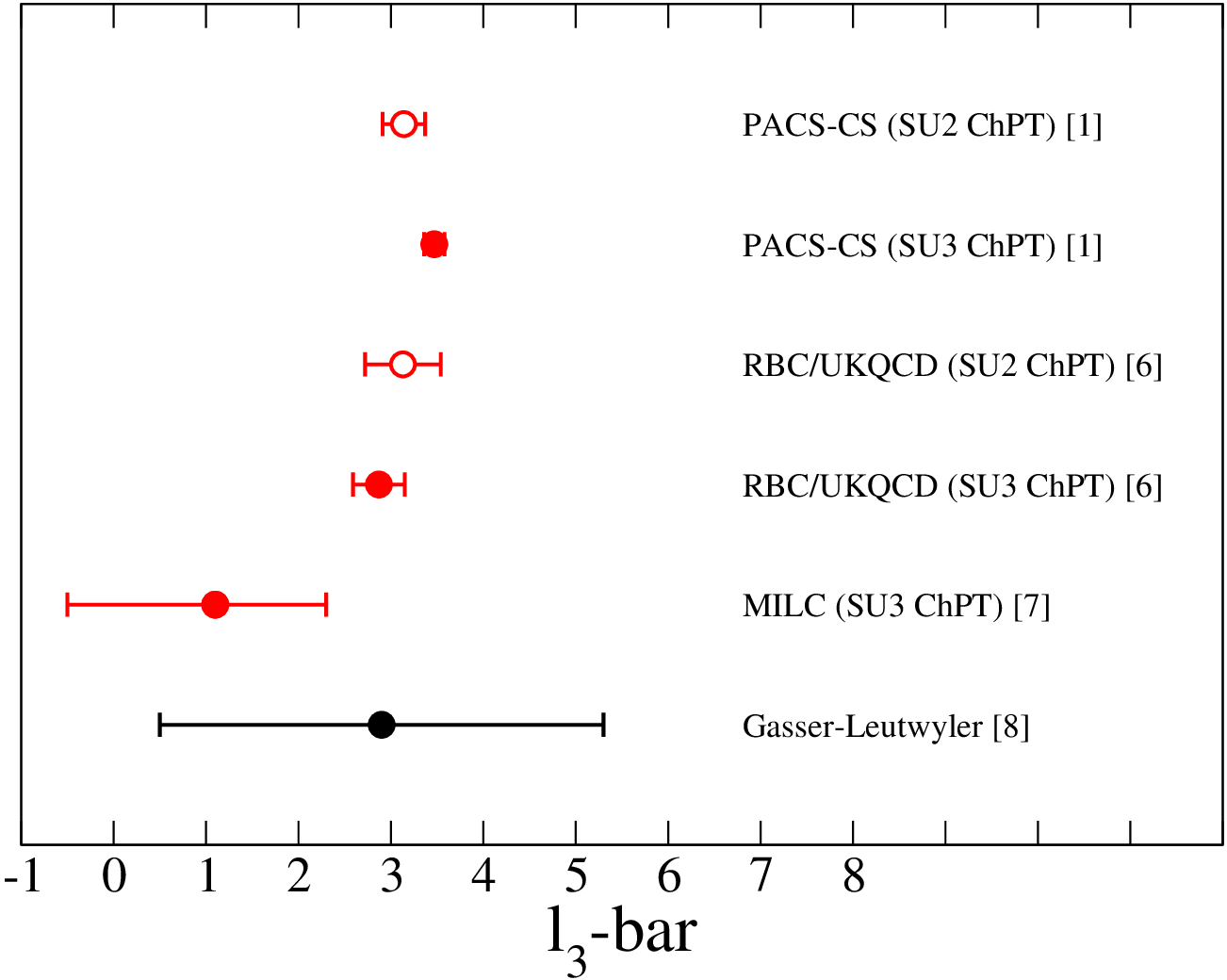,angle=0,height=50mm}
      \caption{Comparison of ${\bar l}_3$ obtained by
the 2+1 flavor dynamical simulations. See text for details.}
      \label{fig:l3-bar}
    \end{minipage}
  \end{tabular}
\end{figure*}

Before physical point simulations, 
we should examine the logarithmic quark mass dependence 
in the pseudoscalar meson sector predicted by the Chiral Perturbation 
Theory (ChPT). 
This is a good testing ground to check whether or not 
the light quark simulations are properly performed. 
In Fig.~\ref{fig:mpi2mud} we plot the ratio $m_\pi^2/m_{\rm ud}$
as a function of $m_{\rm ud}$ in lattice unit  together with the
previous CP-PACS/JLQCD results for comparison. 
The curvature observed near
the chiral limit is explained by the SU(2) ChPT prediction:
\ben
\frac{m_\pi^2}{2 m_{\rm ud}}&=& B\left\{
1+\frac{1}{16 \pi^2}\frac{2 m_{\rm ud} B}{f^2}
\ln\left(\frac{2 m_{\rm ud} B}{\mu^2} \right)+4\frac{2m_{\rm ud} B}{f^2}l_3\right\},
\label{eq:mp2mud}
\een
where $B,f,l_3$ are the low energy constants and $\mu$ is the
renormalization scale. 
Figure~\ref{fig:l3-bar} compares our results for ${\bar l}_3$, which is
defined by ${\bar l}_3=-64\pi^2 l_3$ at $\mu=m_\pi$, 
with currently available data given by other groups.\cite{rbcukqcd08,milc07}
Black symbol denotes the phenomenological estimate.\cite{chpt_nf2}  
Red closed (open) symbols are for the results obtained by the SU(3)
(SU(2)) ChPT fit on 2+1 flavor dynamical configurations.
All the results for ${\bar l}_3$ reside between 3.0 and 3.5, 
except for the MILC result which is sizably smaller and marginally 
consistent with others within a large error.\\
\vspace*{1mm}

\vspace*{-2mm}
\section{Why is the Physical Point Simulation Necessary?} 

\begin{figure*}[h]
  \begin{tabular}{lr}
    \begin{minipage}[c]{0.45\textwidth}
\psfig{figure=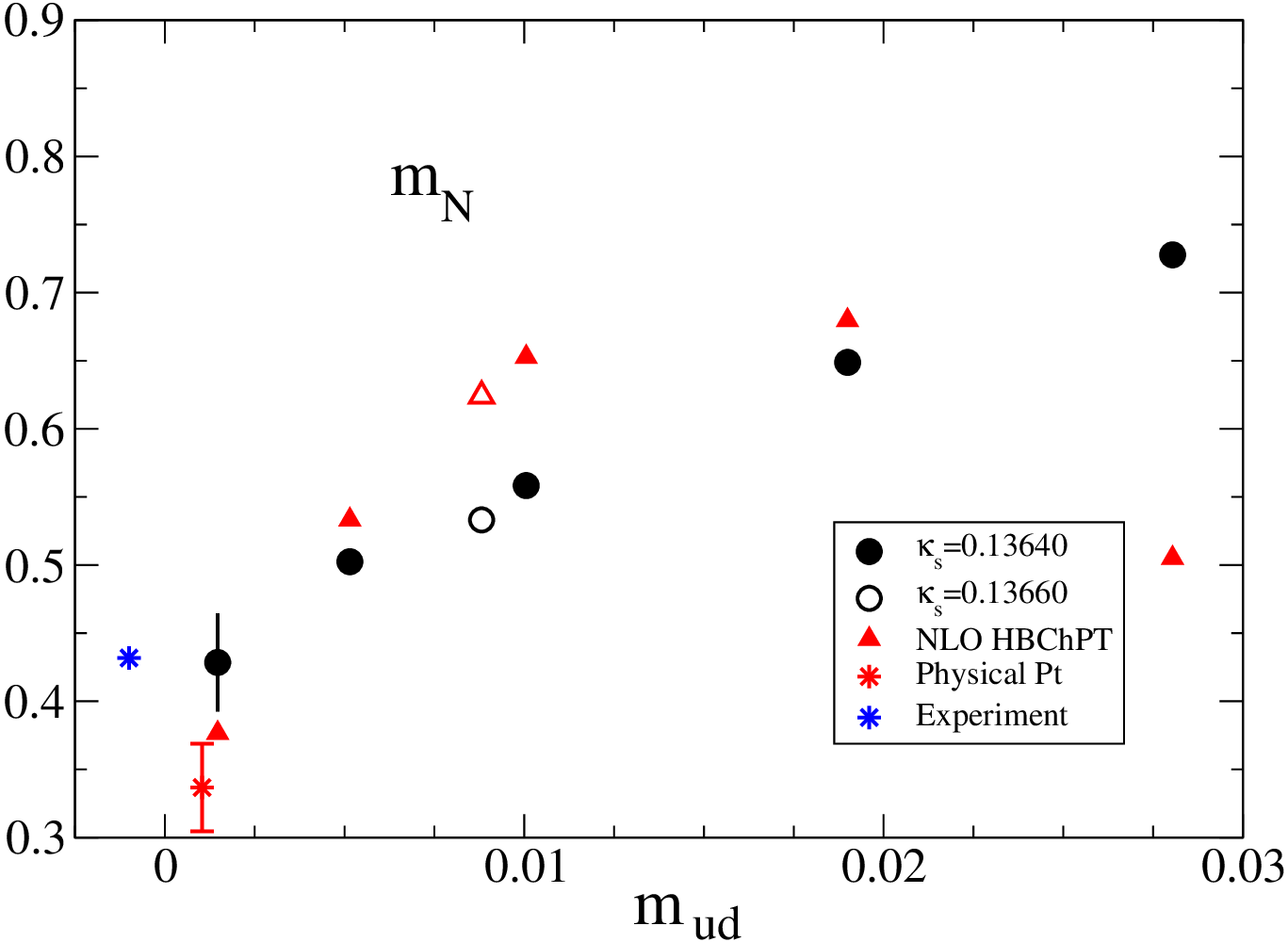,angle=0,height=49mm}
      \caption{Nucleon result of SU(3) Heavy Baryon ChPT fit up to NLO for the
octet baron masses.} 
      \label{fig:mN}
    \end{minipage}
&
    \begin{minipage}[c]{0.45\textwidth}
\psfig{figure=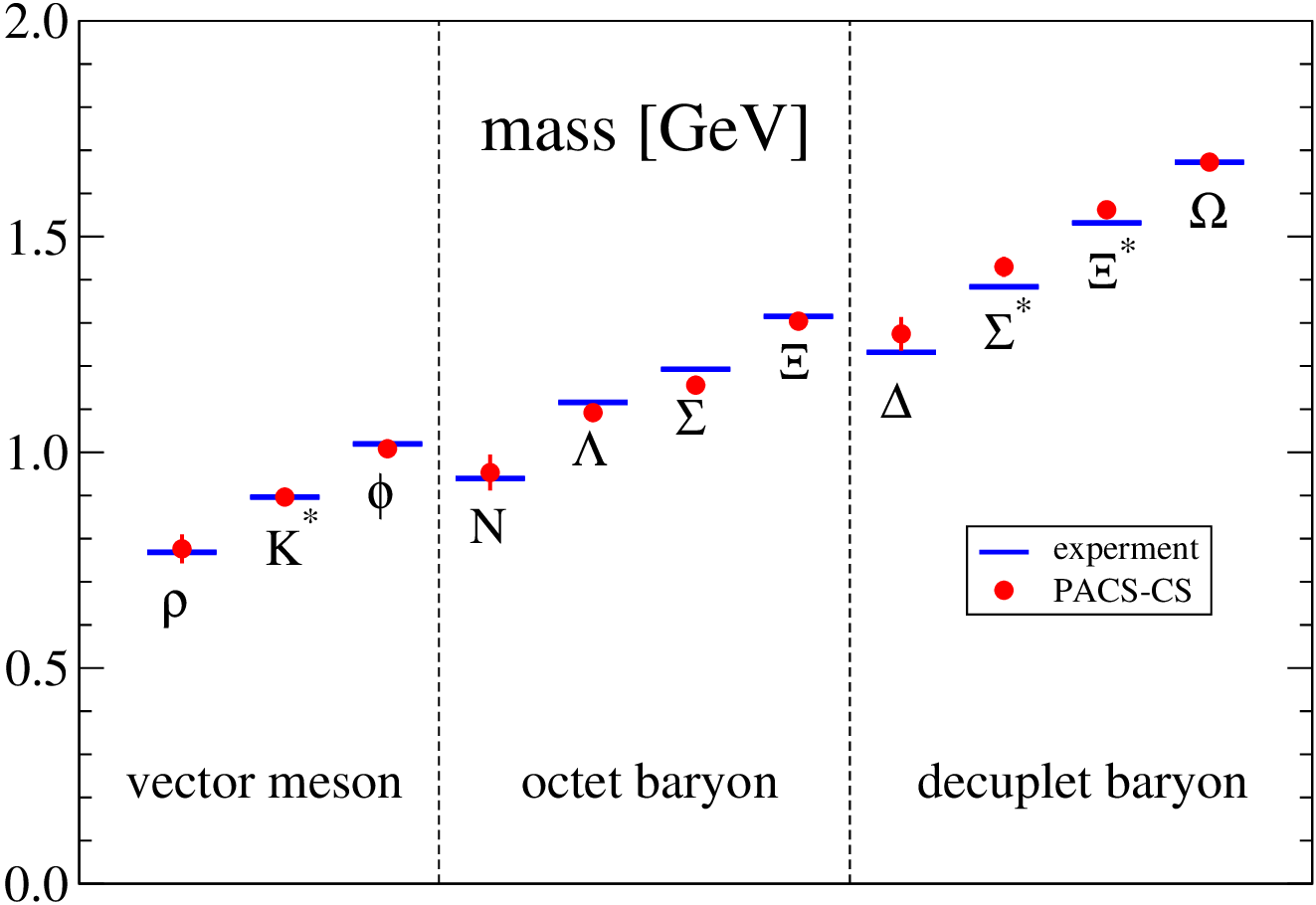,angle=0,height=45mm}
\vspace*{1mm}
      \caption{Light hadron spectrum in 2+1 flavor QCD with 
      $m_\pi$, $m_K$, $m_\Omega$ as physical inputs.}
      \label{fig:spectrum}
    \end{minipage}
  \end{tabular}
\end{figure*}

Chiral extrapolation with the use of ChPT as a guiding principle is
current most popular strategy to estimate the results at the physical point.
The simulation points are usually ranging from $200-300$ MeV to
$600-700$ MeV for $m_\pi$.
There are several problems in this procedure. 
Firstly, it is numerically difficult to trace the logarithmic quark
mass dependence of the physical quantities predicted by ChPT. 
High precision measurements are required for the reliable extrapolation.
Secondly, it is not always possible to resort to the ChPT analyses. 
A typical example is SU(3) Heavy Baryon ChPT 
which completely fails to describe the 
lattice results for the octet baryon masses.\cite{pacscs_hbchpt}
Figure~\ref{fig:mN} shows the next-to-leading order (NLO) 
fit result for the nucleon mass. 
This difficulty may be practically avoided 
by the use of the polynomial fit function instead of ChPT. 
We apply a simple linear function of 
$m^{\rm H}=m_0^{\rm H}+\alpha m_{\rm ud}+\beta m_{\rm s}$
to the lattice data obtained at 156 MeV $\le m_\pi \le$ 410 MeV. 
In Fig.~\ref{fig:spectrum} we compare our results for the hadron
spectrum with the experimental values. Most of them are consistent
within the error bars, though some cases show $2-3$\% deviations 
at most. Note that we are left with the $O((\Lambda_{\rm QCD}\cdot a)^2)$ finite 
lattice spacing effects thanks to the nonperturbative 
$O(a)$-improvement employed 
in our formulation.\footnote{Similar hadron spectrum is obtained by 
the BMW Collaboration.\cite{bmw} It is likely that 
their continuum extrapolation using
the simulations at three lattice spacings succeeds in removing     
the $O(\Lambda_{\rm QCD}\cdot a)$ errors which are the leading finite lattice 
spacing effects in their formulation.}
This encouraging result, however, does not mean 
the polynomial extrapolation is a sufficient solution.
Since we know that $m_{\rm ud}=0$ is a singular point in ChPT,
the convergence radius of the analytic expansion 
around the physical ud quark mass is just
$0< m_{\rm ud} < 2 m_{\rm ud}^{\rm physical}$, which roughly
corresponds to 0 MeV $< m_\pi < 190$ MeV.
Thirdly, it is impossible  to make a proper 
treatment of resonances, {\it e.g.}
$\rho$ meson, with the extrapolation method. 
The reason is quite simple: Lattice QCD calculation shows that 
the pion mass quickly becomes heavier as the ud quark mass increases so that
the kinematical condition $2 m_\pi < m_\rho$ is not satisfied
anymore at the unphysically large ud quark mass.
It is theoretically difficult to predict the real world, where
the $\rho \rightarrow \pi\pi$ decay is allowed, by the chiral
extrapolation from the virtual world with the decay forbidden. 
Fourthly, our final destination is to simulate the different up and
down quark masses, which is an essential ingredient for the proton mass
calculation. The isospin breaking effects are so tiny that
the reliable evaluation would be difficult with the chiral
extrapolation method.\\

\vspace*{-2mm}

\section{Conclusion and Future Plan}

In the history of lattice QCD simulation
the computational cost has been a big issue 
preventing us from the direct simulation at the physical point. 
The chiral extrapolation method, which is just
a compromise due to the lack of computational power, has  
intrinsic problems to be avoided.
Thanks to the rapid increase of computational power and the 
algorithmic improvements in last decade the PACS-CS Collaboration 
are now able to simulate the physical point directly on a (6 fm$)^3$. 
We are going to incorporate the up-down quark mass difference 
and the electromagnetic interactions in the next step.

\section*{Acknowledgments}
Numerical calculations for the present work have been carried out
on the PACS-CS computer
under the ``Interdisciplinary Computational Science Program'' of
Center for Computational Sciences, University of Tsukuba.
This work is supported in part by Grants-in-Aid for Scientific Research
from the Ministry of Education, Culture, Sports, Science and Technology
(Nos.
16740147,   
17340066,   
18104005,   
18540250,   
18740130,   
19740134,   
20340047,   
20540248,   
20740123,   
20740139    
).


\section*{References}
\begin{thebibliography}{99}

\bibitem{pacscs_nf3}
PACS-CS Collaboration, S.~Aoki {\it et al.},
\Journal{\PRD}{79}{034503}{2009}

\bibitem{Parisi}
H.~Hamber and G.~Parisi,
\Journal{\PRL}{47}{1792}{1981}.

\bibitem{cppacs_nf0}
CP-PACS Collaboration, S.~Aoki {\it et al.},
\Journal{\PRL}{84}{238}{2000}; \Journal{\PRD}{67}{034503}{2003}.

\bibitem{berlinwall}
A.~Ukawa, {\em Nucl.~Phys.}~B (Proc. Suppl.) {\bf 106}, 195 (2002).


\bibitem{ddhmc}
M.~L{\"u}scher, {\em J.~High Energy Phys.} {\bf 05}, 052 (2003);
{\em Comput. Phys. Commun.} {\bf 165}, 199 (2005).

\bibitem{rbcukqcd08}
RBC and UKQCD Collaborations, C.~Allton {\it et al.}, \Journal{\PRD}{78}{114509}{2008}.

\bibitem{milc07}
C.~Bernard {\it et al.}, {\em Proc. Sci.} LAT2007, 090 (2007).


\bibitem{chpt_nf2}
J.~Gasser and H.~Leutwyler,  {\em Ann. of Phys.} {\bf 158}, 142 (1984).


\bibitem{pacscs_hbchpt}
PACS-CS Collaboration, K.-I.~Ishikawa {\it et al.},
arXiv:0905.0962 [hep-lat].

\bibitem{bmw}
Z.~Fodor, {\em these proceedings}.





\end{thebibliography}
\end{document}